
%
%
\magnification=\magstep1
\hoffset=0.1truecm
\voffset=0.1truecm
\vsize=23.0truecm
\hsize=16.25truecm
\parskip=0.2truecm
\def\newpage{\vfill\eject}

\def\pp{\parshape 2 0.0truecm 16.25truecm 2truecm 13.25truecm}

\def\fun#1#2{\lower3.6pt\vbox{\baselineskip0pt\lineskip.9pt
  \ialign{$\mathsurround=0pt#1\hfil##\hfil$\crcr#2\crcr\sim\crcr}}}
\def\map{\sigma}

\def\sigwig{ {\widetilde \Sigma} }

\def\remap{\sigma_m}
\def\reduce{\Sigma}
\def\rr{ | {\bf x} - {\bf y} | }

\def\diam{ {\cal D} }
\def\fredarrow{ \longrightarrow }

\def\mapdown#1{\Bigg\downarrow \rlap{ $\vcenter{\hbox{$#1$}}$ } }

\def\norm{ {1 \over \langle \Sigma \rangle} }
\def\omap{\map_{\rm obs} }
\def\rmap{\map_R }
\def\delmap{ \Delta \map }
\def\bound{ \epsilon }
\def\newsig{ {\widetilde \Sigma} }
\def\std{ {\tilde \sigma} }
\def\rhobar{ {\langle \rho \rangle} }
\def\mapbar{ {\langle \sigma \rangle} }
\def\x2{ {\bf x}_2 }

%
%
\centerline{}

\bigskip
\bigskip
\centerline{\bf FORMAL RESULTS REGARDING METRIC-SPACE TECHNIQUES}
\centerline{\bf FOR THE STUDY OF ASTROPHYSICAL MAPS}
\vskip 0.25in
\centerline{\bf Fred C. Adams$^1$ and Jennifer J. Wiseman$^2$}
\vskip 0.15in
\centerline{\it $^1$Physics Department, University of Michigan}
\centerline{\it Ann Arbor, MI 48109}
\vskip 0.15in
\centerline{\it $^2$Harvard-Smithsonian Center for Astrophysics}
\centerline{\it 60 Garden Street, Cambridge, MA 02138}
\vskip 0.4in

\centerline{\it submitted to the Astrophysical Journal: 17 September 1993}
\centerline{\it revised: 1 April 1994}

\vskip 0.4in

\centerline {\bf Abstract}

We extend a newly developed formal system for the description
of astrophysical maps.  In this formalism, we consider the
difference between maps to be the distance between elements
of a pseudometric space (the space of all such maps).
This ansatz allows us to measure quantitatively the
difference between any two maps and to order the space of
all maps. For each physical characteristic of interest, this
technique assigns an ``output'' function to each map; the
difference between the maps is then determined from the
difference between their corresponding output functions.
In this present study, we show that the results of this
procedure are invariant under a class of transformations of
the maps and the domains of the maps.  In addition, we study the
propagation of errors (observational uncertainties) through this
formalism.  We show that the uncertainties in the output functions
can be controlled provided that the signal to noise ratios in the
original astrophysical maps are sufficiently high.  The results
of this paper thus increase the effectiveness of this formal
system for the description, classification, and analysis of
astrophysical maps.

\bigskip
\noindent
{\it Subject headings:} methods: analytical -- methods: data analysis

\newpage
\bigskip
\centerline{\bf 1. INTRODUCTION}
\medskip

Although many different types of astrophysical data are found in maps,
the science of form description for these types of data structures
remains poorly developed (see, e.g., Lord \& Wilson 1984).
In a previous paper (Adams 1992; hereafter Paper I),
we presented a mathematical formalism for the analysis
and classification of astrophysical maps.  This formalism
considers the collection of all maps of a given type as a metric
space.  By defining distance functions (pseudometrics) on the space of
maps, we can measure the difference between any two maps in a
quantitative manner.  We can also provide an ordering scheme for the
space of maps by measuring the ``distance'' from each map to a
well-defined reference map; we can then order the elements of the
space (the maps) by the size of these ``distances''.  In Paper I, we
presented the formalism and proved basic results.  However, before
this formal system becomes a fully functional astrophysical technique,
we must show that it can adequately deal with the standard
difficulties associated with astronomical data, e.g., observational
uncertainties, calibration errors, varying distances to sources,
etc. The overall goal of this paper is to
prove a series of results which show how this formalism behaves
in the presence of the aforementioned astronomical difficulties.
Our hope is to place this formal system on a solid theoretical
footing which will allow for its effective use in the analysis
of astrophysical maps.  In a companion paper (Wiseman \& Adams
1994; hereafter WA), we illustrate the efficacy of this approach
by applying it to a collection of continuum maps of molecular clouds;
in particular, we show that this formal system provides
an effective means of describing and classifying molecular clouds.

Previous studies involving form description in astrophysics
have often been concerned with the large scale structure of the
universe (see, e.g., the reviews of Shandarin \& Zel'dovich
1989; Melott 1990). For example, Gott, Melott, \& Dickinson
(1986) have studied large scale structure by calculating the
topological genus of the surfaces separating high density regions
from those of low density.  Elizalde (1987) has presented a
metric which measures the distributions of voids and clusters
for point distributions (such as distributions of galaxies in
the universe).  Another topic that has been explored is the
possibility that fractal structure appears in observed maps of
molecular clouds (e.g., Bazell \& D\'esert 1988;  Dickman, Horvath,
\& Margulis 1990; Falgarone, Phillips, \& Walker 1991).
Clumps and clump mass spectra in molecular clouds have been
studied over the last decade (e.g., Larson 1981, 1985;
Williams, de Geus, \& Blitz 1994).
Recently, more detailed studies of substructure within molecular clouds
have begun (Myers 1991; Houlahan \& Scalo 1992; Scalo 1990;
Veeraraghavan \& Fuller 1991; Wood, Myers, \& Daugherty 1994;
see also our companion paper).  However, as emphasized by
Elmegreen (1993) in his  recent review, a well defined mathematical
framework with which to describe molecular clouds (and other
astrophysical maps) is badly needed.

The fundamental concept involved in the formal system of Paper I
is that we can consider astrophysical maps as elements of a metric
space; the difference between any two maps is then the ``distance''
between two elements of a metric space (the space of all such maps).
We measure this ``distance'' by constructing a distance function
(a metric), written as
$$d (\sigma_A ,\ \sigma_B ) , \eqno(1.1) $$
where $\sigma_A ,\ \sigma_B$ are maps.  As discussed in Paper I, these
distance functions are actually {\it pseudometrics} and they make the
space of all maps into a {\it pseudometric space.}  The basic
difficulty in this procedure is that the maps are often very
complicated structures and hence the interpretation of any distance
function becomes complicated.  We thus want to simplify our
interpretation of distance.  We therefore ``simplify'' each map
$\sigma$ by ``measuring'' some physical characteristic of the
map.  For example, we can easily determine what fraction of the map
has values above a certain threshold level $\reduce$;
we thus obtain this fraction as a function of the
threshold level.  If we interpret the maps as a measure of
density (or column density), we obtain a function which
represents the distribution of density in the map.  We denote this
profile for a given map as $m (\map; \reduce)$.  We then measure the
difference between any two maps by measuring the distance between
their profiles of density, i.e., we find
$$d \, [ m (\sigma_A ),\ m (\sigma_B ) ] \ . \eqno(1.2)$$
The distance as given by equation (1.2) now has a straightforward
physical interpretation:  the distance measures the difference between
the distributions of density of the two maps.  In addition to the
distribution of density, we can construct other profiles
(as a function of threshold level) which measure other properties
of the maps, e.g., the number of parts the maps break up into,
the shape of these parts, the self-gravity of the maps, etc.
These profiles (usually one dimensional functions of the threshold
level) are denoted as {\it output functions.}  The essential feature
of this formalism is that all of the above concepts can be made
mathematically precise.  We can thus obtain quantitatively
meaningful results.

In this paper, we extend our formal system for the study of astrophysical
maps in two different ways: (1) We prove a series of results which
show how the output functions (and hence the distance functions) vary
when we make various transformations of the original maps.  These
transformations include scalings of the map values, rotations,
translations, stretching the maps, and scalings of the map sizes.
We also briefly discuss resolution and projection effects.
(2) We study the propagation of observational uncertainties
through this formalism.  In particular, we show how observational
uncertainties in the original maps produce corresponding
uncertainties in the output functions.  We find that the errors
can be controlled provided that the signal to noise ratio of
the original map is sufficiently large.

This paper is organized as follows.  In \S 2,  we review the
formal system for measuring the distance between astrophysical maps
and ordering the space of all maps.  We then extend this formalism
in \S 3, where we prove several new results which quantify the manner
in which transformations of the maps affect the output functions.
In \S 4, we discuss the effects of
random errors in the maps and other observational effects.
We conclude in \S 5 with a discussion and summary of our results.
The mathematical details of the effects of transformations and
observational uncertainties on the formalism are presented in
the appendices.

\newpage
\bigskip
\bigskip
\centerline{\bf 2. REVIEW OF THE FORMALISM}
\medskip

In this section, we review the formalism of Paper I.  As discussed
above, we consider each astrophysical map to be an element of an
abstract space -- the space of all possible maps (Paper I; see
also Elizalde 1987).  We thus consider the difference between two
maps to be the ``distance'' between elements of this space.
The space of all maps is thus a metric space (actually a pseudometric
space) and our goal is to define distance functions (pseudometrics)
on the space. \footnote{$^\dagger$}{The difference between a metric
space and a pseudometric space is that the latter can have two
distinct elements with zero distance between them. A distance
function that allows two distinct elements to be separated by
zero distance is known as a pseudometric.  In the study of
astrophysical maps, we want to allow two different maps to be
``the same'' and hence we require a pseudometric space
(see Paper I for further discussion). }

The procedure we use to define distance functions on the space
of all maps contains two steps.  For a given physical characteristic
of interest, we first determine a one-dimensional function (denoted
here as an {\it output function}) to each map; this output function
represents a profile of some physically meaningful quantity (e.g.,
distribution of density -- see below).  For the second step of
this procedure, we determine the difference between any two maps
by finding the difference between their corresponding output functions.
This difference, in turn, is measured using a standard distance
function (denoted here as $d$) defined on the space of functions
(for further discussion of standard distance functions and metric
spaces, see, e.g., Copson 1968; Kelley 1955; Edgar 1990).

This formal system also allows for the ordering of a collection of
maps.  This ordering is accomplished by assigning a positive
real number --  a {\it coordinate} -- to each map, where the
coordinate is defined to be the distance between the map and a
well-defined reference state (or set of states).

This entire procedure can be depicted schematically as:
$$\matrix{
\, & \, & X = \Bigl\{ \map \, \big| \, \map \, \, {\rm is}
\, \, {\rm a} \, \, {\rm map} \Bigr\} &
\buildrel \chi \over \fredarrow & Y_\chi \cr
\, & \, & \, & \, & \, \cr
\, & \, & \mapdown {d \circ \chi} & \, & \mapdown d \cr
\, & \, & \, & \, & \, \cr
\, & \, & \Bigl( X, d \circ \chi \Bigr) &
\buildrel I \over \fredarrow &
\Bigl( Y_\chi, d \Bigr) \cr
\, & \, & \mapdown {d \circ \chi |_{\map_0} } & \, & \, \cr
\, & \, & \, \bigl\{ {\rm coordinates} \bigr\} \subset
{\bf R}^+ & \, & \, \cr
} $$
In the above diagram, we have used the symbol $\chi$ to represent the
assignment of an output function to a given map.  The space $Y_\chi$
is the space of output functions (one output function for each
astrophysical map in the space $X$ of all maps).
We make the space of output functions into a metric space
$(Y_\chi, d)$ by using a standard metric $d$ (as we discuss
below, we generally take $d$ to be the usual $L_2$ norm).  However,
we can assign a distance function $d \circ \chi$ (where $\circ$
denotes composition) directly on the
original space of maps to produce a pseudometric space
$(X, d \circ \chi)$.  The two spaces are related through a
distance preserving function $I$ known as an isometry.
Finally, we assign coordinates to the maps through the
operation denoted as $d \circ \chi \big|_{\map_0}$, which
measures the distance from the map to the nearest reference
map $\map_0$ (see \S 2.4).  Further details of this procedure
are discussed in Paper I. In the following subsections, we describe
some of the output functions that can be used for the description
and study of astrophysical maps.

\medskip
\centerline{\it 2.1 Distributions of Density and Volume}
\medskip

As mentioned above, one approach to characterizing a map to
determine how much of the material is at the highest densities
[We note that the map need not be a density map -- we use this
interpretation only as a conceptual guide.  In most astrophysical
applications, the maps will be of either column density or intensity.]

We can define the fraction of the material at high densities
in two different ways.  We first determine the fraction $m$
of the mass in the map at densities higher than a
given reference $\reduce$:
$$m (\map; \reduce) \equiv
{ \int d^n {\bf x} \, \, \map( {\bf x}) \, \,
\Theta \bigl[ \map({\bf x}) - \reduce \bigr] \over
\int d^n {\bf x} \, \, \map ({\bf x}) } ,  \eqno(2.1)$$
where $\Theta$ is a step function
and where the integrals are taken over the (bounded)
domain $D$ of the map.  Notice that, for a given map $\map$, $m$ is
a function of one variable (namely $\reduce$).  We can also define
an analogous function $v(\map; \reduce)$ which measures the fraction
of the volume (area in a 2-dimensional map) greater than the reference
density $\reduce$:
$$v (\map; \reduce) \equiv
{ \int d^n {\bf x} \, \, \Theta \bigl[ \map({\bf x}) - \reduce \bigr]
\over \int d^n {\bf x} \, \, } .  \eqno(2.2)$$
Given these definitions, we can define a distance between two maps
by measuring the difference between their corresponding output
functions (using either $m$ or $v$), i.e., we
define a pseudometrics $d_m$ and $d_v$ through
$$d_m (\map_A, \map_B) = \Biggr[ \norm \int_0^\infty  d \reduce \,
\big| m (\map_A; \reduce) - m (\map_B; \reduce)
\big|^2 \Biggr]^{1/2} \, ,  \eqno(2.3)$$
$$d_v (\map_A, \map_B) = \Biggr[ \norm \int_0^\infty  d \reduce \,
\big| v (\map_A; \reduce) - v (\map_B; \reduce)
\big|^2 \Biggr]^{1/2} . \eqno(2.4)$$
Notice that in all cases, both $m = 0$ and $v = 0$ at sufficiently
large threshold densities $\reduce$; the integrals in the above
equations are thus always convergent.  Notice also that we have
made the metrics non-dimensional by normalizing the integrals
with an appropriate reference density $\langle \Sigma \rangle$.

The output functions $m(\sigma; \reduce)$ and $v(\map; \reduce)$
have another useful interpretation.  Let us define ${\cal P}_m$
to be (minus) the derivative of the function $m$ with respect to
the variable $\reduce$, i.e.,
$${\cal P}_m (\map; \reduce) = - {d m \over d \reduce} =
{ \int d^n {\bf x} \, \, \map ({\bf x}) \delta
\bigl[ \map({\bf x}) - \reduce \bigr] \over
\int d^n {\bf x} \, \, \map ({\bf x}) } \, ,  \eqno(2.5)$$
where we have used the fact that the derivative of a step
function $\Theta$ is a delta function $\delta ({\bf x})$.
Similarly, we define ${\cal P}_v$ via
$${\cal P}_v (\map; \reduce) = - {d v \over d \reduce} =
{ \int d^n {\bf x} \, \, \delta
\bigl[ \map({\bf x}) - \reduce \bigr] \over
\int d^n {\bf x} \, \, } \, .  \eqno(2.6)$$
The quantity ${\cal P}_v$ is the probability (per unit column density)
of a point in the map $\map$ having the column density $\reduce$.
Similarly, the quantity ${\cal P}_m$ is the probability (weighted
by the mass) of a point in the map $\map$ having the column density
$\reduce$.  It is straightforward to show
that these probability functions are properly normalized, i.e.,
$\int {\cal P}_v d\reduce$ = 1 and $\int {\cal P}_m d\reduce$ = 1.
The interpretation of the derivatives of $m$ and $v$ as probability
distributions greatly facilitates our understanding of how these
output functions behave under various transformations (see below).

\medskip
\centerline{\it 2.2 Distribution of Components}
\medskip

We now consider a diagnostic which can discriminate between different
geometrical distributions of the high density material.  For example,
a given map could be composed of $N$ separate ``units'', each with mass $M$
and a given distribution of density; alternately, the map could consist
of a single unit (with mass $N M$) with the same distribution of density.
Both maps would have the same profile $m(\reduce)$ as defined
above, but would be quite different.  One way to discriminate between
these two cases is to count the number of topological components
as a function of threshold density $\reduce$.  To be
precise, we first define a reduced space:
$$X^+_\reduce \equiv \Bigl\{ {\bf x} \in D \, \Big| \,
\map ( {\bf x} ) > \reduce \Bigr\} . \eqno(2.7)$$
For a given threshold density, the space
$(X^+_\reduce, d_E)$ has a well defined number
$n(\map; \reduce)$ of topological components (where
$d_E$ is the usual Euclidean metric).  We can then define
a pseudometric $d_n$ on the space $X$ of all maps through
$$d_n (\map_A, \map_B) = \Biggr[ \norm \int_0^\infty  d \reduce \,
\big| n (\map_A; \reduce) - n (\map_B; \reduce)
\big|^2 \Biggr]^{1/2} . \eqno(2.8)$$
Since $\map ({\bf x})$ is always a bounded function
$X^+_\reduce = \emptyset$ for sufficiently large $\reduce$;
thus, $n = 0$ at sufficiently large values of $\reduce$ and
the integral is convergent.

Notice that this pseudometric provides us with topological
information on two different levels. For a given threshold
density $\reduce$, the function $n(\map; \reduce)$
measures a topological property (the number of components)
of the {\it reduced space} $X^+_\reduce$, which is in turn
derived from the original map $\map$.  When combined
with our ``standard'' function space metric $d$, the component
function $n$ defines a pseudometric
(namely $d_n = d \circ n$) on the space $X$ of all maps.

\medskip
\centerline{\it 2.3  Distribution of Filaments}
\medskip

We also require some description which measures the shapes of
the pieces of the map. Given the breakup of a map into components
(as described above), we can obtain a measure of the degree to which
the components are filamentary (i.e., stringlike).  We begin with the
usual definition of the diameter $\diam$ of a set $A$, i.e.,
$$\diam (A) \equiv {\rm max} \Bigl\{ \big| {\bf x} - {\bf y}
\big| \, \, \Big| \, \, {\bf x} , {\bf y} \in A
\Bigr\} . \eqno(2.9)$$
For a given threshold density, an astrophysical map breaks up
into components as described in the previous section; each of
these components has a well defined diameter.  We can also
calculate the area ${\cal A}$ of a given component.  Notice
that for a perfectly round (circular) component, the area and
the diameter are related by the obvious relation $\cal A$ =
$\pi \diam^2/4$.  In order to obtain a measure of the
departure of a given component from a circular shape, we
first define a factor ${\cal F}_j$, which is simply the
inverse of the filling factor for a given component, i.e.,
$${\cal F}_j \equiv { \pi \diam_j^2 \over 4 {\cal A}_j } \, .
\eqno(2.10)$$
We denote the quantity ${\cal F}_j$ as the ``filament index''
of the $jth$ component. We also define an average factor $f$:
$$f (\map; \reduce) = {1 \over n(\map; \reduce) }
\sum_j \, w_j \, {\cal F}_j , \eqno(2.11)$$
where the sum is taken over all of the components and where
$f$ is  explicitly written as a function of threshold density
$\reduce$.  The quantities $w_j$ are weighting values; we
consider both an unweighted version of the filament index
($w_j$ = 1) and a weighted version in which each ${\cal F}_j$
is weighted by the fraction of material in that component
(i.e., $w_j \equiv A_j / \langle A \rangle$, where
$\langle A \rangle$ is the average area of the components at
the given threshold level).
A highly filamentary map will thus have a very large value of $f$.
The pseudometric $d_f$ on the space of maps then can be written
$$d_f (\map_A, \map_B) = \Biggr[ \norm \int_0^\infty  d \reduce \,
\big| f (\map_A; \reduce) - f (\map_B; \reduce)
\big|^2 \Biggr]^{1/2} \, ,  \eqno(2.12)$$
where $f$ can be either the weighted or unweighted version of
the filament function [2.11].

\medskip
\centerline{\it 2.4 Assigning Coordinates}
\medskip

In applications, we often require a method of putting
astrophysical maps into some kind of well defined order.
To order the space of all maps using this formal system,
we assign ``coordinates''
to the maps by measuring the distance from a given map $\map$
to a well-defined reference map $\map_0$. The resulting
coordinates are thus positive real numbers and the space
of maps can be ordered by the size of these numbers.
Notice that, in general, there will be different coordinates
(and hence a different ordering) for each type of output
function considered.

As described in Paper I, we use uniform density maps
($\map_0$ = {\it constant}) as reference maps.  Notice that
there are an infinite number of such reference maps -- one
for each possible value of the constant.  We follow Paper I
in defining the coordinate to be the distance to the
{\it nearest} uniform density map. Specifically, for a given
pseudometric $d_\chi$, we define the coordinate $\eta_\chi$ by
$$\eta_\chi \equiv {\rm min} \Bigl\{ d_\chi (\map, \map_0)
\Big| \map_0 \, \, {\rm is} \, \, {\rm a} \, \, {\rm uniform}
\, \, {\rm density} \, \, {\rm map} \Bigr\} \, . \eqno(2.13)$$
See Paper I for further details on implementing this
minimization procedure.  We can thus say that map $\map_A$
is ``greater than'' map $\map_B$ provided that
$$\eta_\chi(\map_A) > \eta_\chi(\map_B) \, . \eqno(2.14)$$

In addition to the coordinates obtained from the output functions
described above, we can define other coordinates.  For example,
in applications (see WA) we sometimes want to consider the
``self-gravity'' of a map.  Thus, for a given map we define
the quantity
$$\eta_w \equiv {1 \over 2} \, {\cal N}_w \,
\int d^n {\bf x} \int d^n {\bf y} \, \,
{\map({\bf x}) \map({\bf y}) \over \rr} \, , \eqno(2.15)$$
where $\bf x$ and $\bf y$ are ($n-$dimensional) position vectors
and where ${\cal N}_w$ is a normalization factor which is chosen
to make the coordinate a dimensionless quantity (see Paper I).
The integrals are carried out over the entire map.
The quantity $\eta_w$ thus plays the role of an additional
coordinate.  If the map traces a mass distribution, then
this coordinate provides a measure of the self-gravity of
the distribution.

\bigskip
\bigskip
\centerline{\bf 3. BEHAVIOR OF THE FORMALISM UNDER TRANSFORMATIONS}
\medskip

In this section, we present formal results which show how the
formalism described in the preceding section behaves under
various transformations of the maps.  In particular, we discuss
how the output functions transform under scaling transformations
which change the values of the maps, but not the spatial positions
of the map pixels in the sky.   Next, we study the effects of
changing the domains of the maps for a general class of affine
transformations (e.g., rotations, translations, stretching of
the maps).

\medskip
\centerline{\it 3.1 Scaling Transformations}
\medskip

We now discuss what happens to the output functions under
various types of scaling transformations.  For the sake of
definiteness, we begin with the simplest type of scaling
transformation, i.e.,
$$ \map \to \beta \map , \eqno(3.1)$$
where $\beta$ is a constant.  We will also consider more
general types of rescalings of the maps.

Scaling transformations (e.g., equation [3.1]) can arise in many ways.
For example, overall calibration considerations can necessitate an
overall scaling transformation of this type.  More complicated
scaling transformations arise in practice due to calibration
problems/uncertainties (see the discussion of WA and Theorem 2
below).

Another example of scaling occurs due to the two dimensional nature
of many astrophysical maps (such as the {\it IRAS} maps considered
in our companion paper).
These maps only measure the column density $\map ({\bf x}_2)$, whereas
we would like to know the true volume density $\rho ({\bf x}_3)$, where
we have explicitly used subscripts to refer to the number of spatial
dimensions. We are thus limited to considering only the average
of the volume density over the line of sight.
\footnote{$^\dagger$}{We note that the column density is a measure
of the average density only when the emissivity of the emitting
medium is sufficiently well behaved.  In the case of the {\it IRAS}
maps considered in our companion paper (WA), for example, the
presence of temperature gradients in the maps can lead to
interpretation problems -- see Langer et al. (1989).}
We can conceptually break up the observed map $\map$
into two parts through the ansatz
$$\map(\x2) = \rhobar (\x2) \, L(\x2) , \eqno(3.2) $$
where $L(\x2)$ is the depth of the object (along the line of
sight) at position $\x2$ in the plane of the sky.
In many applications, the interpretation of maps is driven by
the assumption that the observed maps trace the mean density $\rhobar$
and the function  $L(\x2)$ is slowly varying compared to
$\rhobar$. As we show in WA, this assumption can be
justified by the analysis of real data (at least for some
cases). In the sample of WA, the observed maps vary by
several orders of magnitude in $\map$. On the other hand,
the filament index (which measures the shapes in the plane
of the sky -- see \S 2.3) is rarely greater than 3 and is
typically $\sim 2$.   Since we do not expect that
the variation in map size along the line of sight is markedly
different from the observed variations in the plane of the sky,
we expect that the depth function $L(\x2)$ varies only by
factors of 2 or 3 across the face of a typical map.  We can
thus regard $L(\x2)$ as a slowly varying function.
We can now consider the question: What happens when we want
to compare two maps which are essentially the same in all
regards except that one map is twice as large as the other?
If we can take the extreme limit where $L(\x2) = L = $
{\it constant} for each map, then we are comparing two
maps related by a scaling transformation of the form [3.1]
where $\beta$ = $1/L$.

Thus, it is useful to know how the output functions behave
under scaling transformations.  For simple transformations of
the form of equation [3.1], the answer to this question can be
given as follows:

\proclaim Theorem 1. Let $\map$ be any map and let
$\chi(\map; \reduce)$ be any of the following output functions:
distribution of density $m(\map; \reduce)$, volume $v(\map; \reduce)$,
components $n(\map; \reduce)$, or filaments $f(\map; \reduce)$.
Then the output function scales under the transformation
$\map \to \beta \map$ according to
$\chi(\map; \reduce)$ $\to$ $\chi(\map; \reduce/\beta)$.
\par

\noindent
A proof is given in Appendix A. Theorem 1 provides us with a
very useful property of this formal system.  For example, if we
want to consider an alternate calibration of the original
astrophysical maps, we can simply transform the output functions
after the fact and need not recalculate them. As discussed above,
we can also use Theorem 1 to compare maps of different physical
sizes provided that we know the appropriate scaling factor
$\beta$ from independent observations.

Given the dependence of the output functions on the
scaling transformation [3.1], we can determine the dependence
of the coordinates on scaling.  This result can be stated as
the following corollary, which follows directly from the
definition of coordinates and Theorem 1:

\proclaim Corollary. Let $\map$ be any map and let
$\chi(\map; \reduce)$ be any of the following output functions:
$m(\map; \reduce)$, $v(\map; \reduce)$, $n(\map; \reduce)$,
or $f(\map; \reduce)$. Then the coordinate $\eta_\chi$ scales
under the transformation equation [3.1] according to
$\eta_\chi (\beta \map)$ = ${\sqrt \beta} \, \eta_\chi (\map)$.
\par

We now consider a more general type of rescaling
law in which the maps are transformed according to
$$\map \to F(\map) \map , \eqno(3.3)$$
where $F$ is a given function of $\map$.
In order for this transformation to be ``well behaved'',
we must make the further restriction that the quantity
$F(\map) \map$ is a monotonic function of $\map$.
Notice that we have written the transformation
in the form $F(\map)\map$ rather than simply $G(\map)$
for convenience; in the transformations of this type
encountered for {\it IRAS} maps (see WA; Jarrett, Dickman,
\& Herbst 1989), $F = 1$ for small
values of $\map$ and only departs significantly from
unity at large map values.

We now discuss how the output functions behave under this
transformation.  Suppose, for example, that we have already
computed output functions for a sample of maps.  We
may want to consider an alternate calibration procedure
(which will result is a transformation of the maps according
to equation [3.3]). Since the calculation of the output
functions is computationally expensive, we would like to be
able to simply transform the output functions after the fact.

We first note that, for a given threshold level $\reduce$,
the equation
$$F(\map) \map = \reduce  \eqno(3.4)$$
has (at most) one root because of our assumption that the left
hand side is monotonic in $\map$. We denote this root by
$\map = \sigwig$, which is simply a rescaling of the
threshold level under the transformation. Given this definition,
we can determine the manner in which the output functions
transform under equation [3.3].  We state this result as
the following theorem:

\proclaim Theorem 2. Under transformations of the form given
in equation [3.3], the output functions transform according to
$\chi(\map; \reduce)$ $\to$ $\chi(\map; \sigwig)$, where $\chi$
represents any of the four output functions $m(\map; \sigwig)$,
$v(\map; \sigwig)$, $n(\map; \sigwig)$, or $f(\map; \sigwig)$,
and where $\sigwig$ is the root of equation [3.4].
\par

\noindent
A proof is given in Appendix B.  Theorem 2 shows that even
under complicated rescalings of the maps (such rescalings
can easily arise in practice due to calibration problems),
the output functions transform in a simple manner.

\medskip
\centerline{\it 3.2 Transformations of the Domain}
\nobreak
\medskip

In this section, we study the dependence of the output
functions under transformations which change the domain
of the maps.  We concentrate on a general class of transformations
known as affine transformations and briefly consider the effects
of resolution transformations.

We begin with the general definition of affine transformations:

\proclaim Definition. An affine transformation $W$ in two
dimensions is a transformation of the form
$$W \Biggl[ \eqalign{ & x \cr & y  } \Biggr] =
\Biggl[ \eqalign{ & x^\prime \cr & y^\prime  } \Biggr] =
\Biggl[ \eqalign{ a_{1 \, 1} & \, \quad  a_{1 \, 2} \cr
a_{2 \, 1} & \, \quad a_{2 \, 2} } \Biggr]
\, \Biggl[ \eqalign{ & x \cr & y } \Biggr] +
\Biggl[ \eqalign{ & b_1 \cr & b_2  } \Biggr] \, . \,
\eqno(3.5)$$
\par

\noindent
We denote the matrix in the above definition as $A$ and we assume
that this matrix is non-singular, i.e., $\det A \ne 0$. We are also
interested in a particular type of affine transformation known
as a similitude.

\proclaim Definition. A similitude is a special type of
affine transformation in which the matrix $A$ takes the form
$$A = \Biggl[ \eqalign{ r \cos\theta & \, \,  - r \sin\theta \cr
r \sin\theta & \, \, \, \, r \cos\theta } \Biggr] \qquad {\rm or} \qquad
A = \Biggl[ \eqalign{ r \cos\theta & \, \,  \, \, r \sin\theta \cr
r \sin\theta & \, \,  - r \cos\theta } \Biggr] \, .  \,
\eqno(3.6)$$

\noindent
Thus, a similitude corresponds to a rotation (through angle
$\theta$), a scaling of the domain (by the factor $r$),
a translation (by $[b_1, b_2]$), and a possible reflection.
Affine transformations are more general and include the
possibility of stretching the $\hat x$ direction and the
$\hat y$ direction by different amounts (see, e.g.,
Barnsley 1988 for further discussion of affine transformations).

For astrophysical maps $\map$, we think of affine transformations
as changing the domain $D$ of the map.  For example, we generally
do not know precisely the distance to a given object.  If we want to
conceptually ``change the distance'' to an object, the corresponding
change in the map is a similitude (where $r$ is the ratio of the new
distance to the old distance, $\theta$ = 0, $b_1 = 0 = b_2$).
In general, an affine transformation acts on an astrophysical
map according to
$$\Biggl[ \eqalign{ x & \cr y & \cr z= & \map(x,y) } \Biggr]
\, \to \,
\Biggl[ \eqalign{
W \Bigl[ & \eqalign{ x & \cr y & } \Bigr] \cr
z = & \map(x,y) } \Biggr] \, . \eqno(3.7)$$
Given these definitions we can state
the main result of this section:

\proclaim Theorem 3. The output functions $m(\map; \reduce)$,
$v(\map; \reduce)$, and $n(\map; \reduce)$ are invariant under
general affine transformations, whereas the distribution of
filaments $f(\map; \reduce)$ is invariant under similitudes.
\par

\noindent
Although the behavior of the output functions under affine
transformations is relatively easy to understand, it is
somewhat more difficult to actually prove what happens.
In any case, the proof is presented in Appendix C.
Theorem 3 shows that the output functions transform
in a fairly simple manner under trasformations which
are often encountered in astrophysics; this result
thus provides another justification for
using the formalism of Paper I.

\medskip
\centerline{\it 3.3 Resolution Transformations}
\medskip

In this section, we briefly discuss the effects of
resolution on the maps. We begin by defining a
{\it resolution transformation} $\cal R$ according to
$${\cal R} [\map ({\bf x})] = {\cal N} \int d^2 {\bf x}^\prime
\, \exp[ - ({\bf x}^\prime - {\bf x})^2/L^2 ]
\map({\bf x}^\prime) \, , \eqno(3.8)$$
where $\cal N$ is a normalization factor for the transformation.
If we normalize the transformation such that ${\cal R}(1)$ = 1, we
obtain the normalization factor $\cal N$ = 1/$\pi L^2$.  Notice that
we have explicitly written the integral in two spatial dimensions,
since the transformation occurs in practice due to telescope beams
averaging signals within the plane of the sky.
Here, $L$ is the physical length scale corresponding to the
given beam size of the telescope.  Let us also introduce a
constant physical length scale $R$, which corresponds to the
length scale on which the true map $\map$ varies.  We then define
$$\lambda \equiv 2 R^2/L^2 . \eqno(3.9)$$
It is easy to see that in the limit $\lambda \ll 1$, the
integral in equation [3.8] averages out much of the structure
in the map and we are left with the age-old astronomical problem
of loss of information.  In the opposite limit, $\lambda \gg 1$,
we can perform an asymptotic expansion (Appendix D) of the
transformation in order to quantify this loss of information.
In this limit, the leading order change in the map due to a
resolution transformation is given by
$${\cal R} [\map] = \map + {1 \over 2 \lambda} \Delta_2 \map
+ {\cal O} \bigl( \lambda^{-2} \bigr) \, , \eqno(3.10)$$
where $\Delta_2$ is the two-dimensional Laplacian operator
in dimensionless units (see equation [D5]).
Thus, we see explicitly the expected result that, under a
resolution transformation, the peaks of the map become
less pronounced ($\Delta_2 \map < 0$) whereas the low points
(valleys) of the map become filled in ($\Delta_2 \map > 0$).
Thus, the resolution transformation [3.8] leads to a loss
of dynamic range in the map and the size of this effect
is $\sim$ $\langle \Delta_2 \map \rangle / \lambda$.

\bigskip
\bigskip
\centerline{\bf 4. PROPAGATION OF ERRORS}
\medskip

An important issue which must be addressed is the effect of
observational errors on the classification scheme outlined above.
In particular, we must determine the effects of observational errors
on the output functions and on the corresponding coordinates.
For this present discussion we assume that the observed value
$\omap$ of the map at a particular spatial point and the uncertainty
$\delmap$ in the map at that point can be written
$$\omap = \rmap + \delmap , \eqno(4.1)$$
where $\rmap$ is the true value of the map at that point.
The uncertainty $\delmap$ is usually considered to
have a Gaussian distribution.

The goal of this section is to show that the output functions used
here to describe astrophysical maps are sufficiently ``well behaved''
in the presence of uncertainties in the maps.  In particular, we
want to show that
small uncertainties $\delmap$ in the original maps do not lead to
large variations in the resulting output functions.  Fortunately,
by working within the formal system of Paper I, we can provide a
mathematically precise answer to this question (see Appendices
E, F, and G). As we discuss below, the net result of this analysis
is that observational uncertainties can be well controlled within
this formalism provided that the signal to noise ratios in the
original maps are sufficiently high.

For the distribution of density $m(\map; \reduce)$,
it is fairly straightforward to determine the relative
error $\Delta m/ m$ in the output function in terms of
the uncertainty in the map; we find
$${\Delta m \over m} = {1 \over m} \Big|
{dm \over d\reduce} \Big|_\reduce \delmap \, , \eqno(4.2)$$
where we have used the results of Appendix E.  The errors are thus
well controlled: small uncertainties $\delmap$ in the maps lead to
small relative uncertainties $\Delta m / m$ in the output functions.
In principle, the slope $dm/d\reduce$ can be quite large at a
particular value of $\reduce$ and the corresponding error could be
large; in practice, however, the right hand side of equation [4.1]
does not become overly large (see WA).  Notice that we can average
equation [4.2] over all threshold densities $\reduce$ and thus
obtain the average relative error in the output function; we find
$$\Bigg\langle {\Delta m \over m}  \Bigg\rangle =
{\log(SNR) \over SNR } \, , \eqno(4.3)$$
where $SNR$ denotes the signal to noise ratio (which we
have assumed to be uniform over the map).  We thus conclude
that the distribution of density is sufficiently well
behaved provided that the signal to noise ratio in the map
is sufficiently large. The distribution of volume $v(\map; \reduce)$
can be treated similarly.

We now consider possible errors in the distribution of
components $n(\map; \reduce)$. This present discussion is
complicated by the fact that relatively small errors in the
map can, in principle, produce rather large errors in the
number of components.  For example, the distribution of
components $n(\map; \reduce)$ can have extra components due
to erroneous pixels sticking up above the threshold level.
Notice, however, that the opposite does not generally occur
-- we seldom get extraneous components due to a pixel being too
low.  A pixel with an erroneously low value will show up as a
``hole'' in a component but will not, in general, change the
number of components.

We must estimate the probability $P_1$ that a pixel in the map
will be erroneously larger than a given threshold level.
This estimate is given in Appendix F and shows that the
probability $P_1$ is bounded to be quite small; very
roughly we obtain $P_1$ in the range $10^{-2}$ to $10^{-3}$.
Thus, in a map containing $N$ $\times$ $N$ pixels, the number of pixels
erroneously appearing larger than the threshold $\reduce$
is bounded to be less than $N^2 P_1$. For the representative values
$N$ = 400 and $P_1$ = $10^{-2}$ (these values are appropriate
for the maps considered in WA), we  have $N^2 P_1$ = 1600.
Although most of these erroneous pixels are
not expected to produce erroneous components, this number is
uncomfortably large.  In order to control the number of erroneous
components, we should eliminate all single-pixel components
from the function $n(\map; \reduce)$.  The probability
that two adjacent pixels are both erroneously above a
threshold is proportional to $P_1^2$ and the number of (possible)
erroneous components is bounded by $2 N^2 P_1^2$ $\sim$ 32
(for the same values of $N$ and $P_1$ as before). Although
the number of spurious components with two pixels is bounded
to be fairly small, we should adopt the conservative approach and
also eliminate all two-pixel components from consideration.
The smallest components that are kept in the function
$n(\map; \reduce)$ are those with
three pixels and the probability of obtaining a spurious
component is proportional to $P_1^3$ $\sim$ $10^{-6}$
and the number of spurious components is expected to be
less than $\sim 6 N^2 P_1^3$ $\sim$ 0.96 $<$ 1.
We therefore conclude that the errors in the distribution
of components can be well controlled, provided that we
eliminate from consideration all components with only
one or two pixels. We note that in practice the values
of $N$ and $P_1$ may differ from those taken here for
purposes of illustration, although we have been reasonably
conservative here.  The exercise outlined above
should be performed for each application of the formalism
and, in some cases, components with three (or even more)
pixels might need to be removed in order to control errors
in the distribution of components function $n(\map; \reduce)$.

Another issue which must be mentioned is the choice of
the step size of the threshold level $\reduce$ used in
computing the distribution of components output function.
This step size should be related to the noise levels in
the maps used.  If the step size is much smaller than the
RMS noise of the map, then extraneous noise can appear in
the output function itself.  On the other hand, if the
step size is much larger than the RMS noise level, then
information will be lost and we will lose some ability to
detect structure.

For the distribution of filaments $f(\map; \reduce)$, the
relative error is approximately given by
$${\Delta f \over f_R} = {\Delta n \over n_R} \Biggl\{
1 - {f_S \over f_R} \Biggr\} \, , \eqno(4.4)$$
where $f_S$ is the average filament index of the spurious
components and $f_R$ is the average filament index of the
real components (see Appendix G).  Here, $\Delta n/n_R$
is the relative error in the distribution of components function.
The quantity in brackets in equation [4.4] is expected to be
of order unity or smaller (if $f_S \approx f_R$, then the
errors nearly cancel).  The quantity $f_S$ represents the
average shape (filament index) of the spurious components.
Since these components are made up of pixels which are erroneously
on the wrong side of the threshold level, and since the
probability of erroneous pixels occurring is small,
these components are generally made up of a small number
of pixels.  The possible shapes of such components can
be computed and the average filement index for such shapes
is not far from unity.  Thus, the relative error in the
distribution of filaments is well controlled provided
that $\Delta n / n_R$ is small, i.e., provided that the
number of spurious components is small compared to the
number of true components.  As we have argued above, however,
the number of spurious components can be controlled by
considering only those components with three or more pixels;
thus, $\Delta n$ can be made small and hence $\Delta f$ can
also be controlled.

Although we have shown that uncertainties $\Delta \map$
in the maps do not produce unacceptably large uncertainties
in the output functions themselves, we still must show that
the corresponding uncertainties in the coordinates are
sufficiently well behaved.  Fortunately, Theorem 3 of
Paper I shows that when two maps $\map_A$ and $\map_B$
are close together in the space of maps,
$$d_\chi (\map_A, \map_B) < \epsilon , \eqno(4.5)$$
then the corresponding coordinates ($\eta_A$ and $\eta_B$)
are close together in the space of real numbers, i.e.,
$$\big| \eta_A - \eta_B \big| \, < \, \epsilon \, . \eqno(4.6)$$
This result follows directly from the fact that we
have built this formalism using pseudometrics (which are
{\it uniformly continuous} when considered as functions
of the maps -- see Paper I; Copson 1968). Here, we let
$\map_A = \map_{\rm obs}$ denote the observed map, and we let
$\map_B$ $= \map_R$ denote the true map, i.e., the map we would
have if there were no observational uncertainties (see equation
[4.1]). For any given type of output function $\chi$, the results
of the previous analysis  show that the difference between the
output function $\chi_{\rm obs}$ of the observed map and the
output function $\chi_R$ of the true map is small, i.e.,
$$\chi_{\rm obs} = \chi_R + \Delta \chi \, , \eqno(4.7)$$
where $\Delta \chi$ is small compared to $\chi_R$.  Using the
definitions of the pseudometrics (see equations [2.3], [2.4],
[2.8], and [2.12]), we find that the difference between the
true map and the observed map is given by
$$d_\chi (\map_R, \map_{\rm obs}) = \Biggr[ \norm
\int_0^\infty  d \reduce \, \big| \Delta \chi \big|^2
\Biggr]^{1/2} \, = {\cal O} ( | \Delta \chi | ) \, . \eqno(4.8)$$
Using equation [4.8] and Theorem 3 of Paper I, we find that
the coordinates of the true and the observed maps are also
``close together'' in the sense that
$$\big| \eta_R - \eta_{\rm obs} \big| \, \le \,
{\cal O} ( | \Delta \chi | ) \, . \eqno(4.9)$$
We thus conclude that the uncertainties in the coordinates
are sufficiently well behaved in this formalism.

\bigskip
\bigskip
\goodbreak
\centerline{\bf 5. SUMMARY AND DISCUSSION}
\medskip

In this paper, we have extended the formalism for using
metric space techniques to study astrophysical maps.
In particular, we have proved a number of useful results
concerning this method of form description. These results
are summarized below.

Our first result applies to scaling of the output functions due
to the scaling of the original maps.  Such scalings can arise
from calibration considerations.  In addition, if two maps are
similar except for their size, then the corresponding column density
maps should show similar structure except for an overall
scaling factor $\beta$.  If we apply a scaling transformation
$\sigma \to \beta \sigma$ to a given map, the output functions
transform in a relatively simple manner, i.e.,
$\chi(\map; \reduce)$ $\to$ $\chi(\beta \map; \reduce)$ =
$\chi(\map; \reduce/\beta)$.  This transformation thus allows us
to compare maps which are known to have different sizes.
We have also shown that under more general transformations
of the form $\map \to F(\map) \map$, the output functions
transform according to $\chi(\map; \reduce)$ $\to$
$\chi (\map; \sigwig)$, where $\sigwig$ is the root
of equation [3.4].

Our next result shows how the output functions are affected by a
general class of transformations of the domain of the maps.
This class of affine transformations includes rotations, translations,
and an overall stretching of the domain. The output functions for the
distributions of density, volume, and components are all invariant under
general affine transformations.  The distribution of filaments is
invariant under  a restricted class of transformations known as
similitudes, and is not invariant under general affine transformations.

We have found one additional property of this formal system.
For any given astrophysical map, the distribution of density
output function is always greater than the distribution of volume
(we prove this result in Appendix H).  This property of the output
functions is evident for the maps considered in our companion
paper (WA).

We have discussed the propagation of errors in the original maps into
the corresponding errors in the output functions. The distribution of
density function is well behaved in the sense that small map errors
lead to small errors in the output function. The distribution of
volume function behaves similarly.  The distribution of filaments is
well behaved provided that the number of components can be well
determined.  On the other hand, the distribution of components is
potentially more problematic.  The function $n(\map; \reduce)$ can
have considerable errors due to single erroneous pixels sticking up
above the threshold level.  However, we can estimate the probability
of a spurious component arising with $k$ pixels.  This probability
is $\propto$ $P_1^k$ (where $P_1$ is expected to be a small number --
see Appendix F) and thus the probability of obtaining a spurious
component with a large number $k$ of pixels is quite small.
In practice, we expect that by removing from consideration all
components with only one or two pixels, we can adequately control
the errors in the distribution of components $n(\map; \reduce)$.
Finally, we have shown that uncertainties in the coordinates are
small as long as the corresponding uncertainties in the output
functions are small.

This present paper (in conjunction with Paper I) provides
a well-defined formal system for the quantitative description
of astrophysical maps and other similar data structures.
This formal system can be directly applied to many types
of existing astrophysical maps (see, e.g., WA for an
application to continuum maps of molecular clouds).
In addition, this method of form description can have
applications outside the field of astrophysics. For example,
studies of computer vision and artificial intelligence
require the construction of explicit and meaningful
descriptions of physical objects from images (see, e.g.,
Ballard \& Brown 1982).

In addition to direct applications of this formal system,
many possibilities for future
developmental work remain.  So far, we have only used the most
basic physical quantities (e.g., fraction of high density
material by mass and by volume, number of pieces, and shape
of the pieces) as the basis for our output functions. As the
study of astrophysical form description becomes better
understood, additional output functions should be developed
to provide a better description of these maps. These additional
output functions should be constructed so that they are
either invariant under the transformations discussed in \S 3
or they transform in a simple manner. These new output functions
should also be ``well behaved'' in terms of their sensitivity to
observational uncertainties in the maps.

\vskip 0.6truein
\centerline{Acknowledgements}
\medskip

We would like to thank Brian Boonstra, Dick Canary, Marco Fatuzzo,
Paul Ho, Phil Myers, Greg Thorson, Rick Watkins, and Doug Wood for
helpful comments and useful discussions.  We would also like to thank
the referee -- Eugene de Geus -- for many useful comments and
criticisms. This work was supported by  NASA Grant Nos. NAGW--2802
and NAGW--3121, the NSF Young Investigator Program, and by funds
from the Physics Department at the University of Michigan.

\bigskip
\bigskip
\bigskip
\goodbreak
\centerline{\bf APPENDIX A: PROOF OF THEOREM 1}
\medskip

In this Appendix, we present a proof of Theorem 1, which
states that the output functions scale under the transformation
$\map \to \beta \map$ according to $\chi(\map; \reduce)$ $\to$
$\chi(\map; \reduce/\beta)$, where $\chi(\map; \reduce)$ represents
any of the following output functions: distribution of density
$m(\map; \reduce)$, volume $v(\map; \reduce)$, components
$n(\map; \reduce)$, or filaments $f(\map; \reduce)$.

\noindent
{\it Proof:}
For the distribution of density $m(\map; \reduce)$, we
begin with the definition of $m$, i.e.,
$$m(\map; \reduce) = { \int d^n {\bf x} \, \,
\map ({\bf x}) \, \Theta \bigl[ \map({\bf x}) - \reduce \bigr]
\over \int d^n {\bf x} \, \, \map ({\bf x}) }  .
\eqno({\rm A}1)$$
Under the transformation [3.1] we obtain
$$m(\beta \map; \reduce) = { \int d^n {\bf x} \, \, \map
({\bf x}) \, \Theta \bigl[ \beta \map({\bf x}) - \reduce \bigr]
\over \int d^n {\bf x} \, \, \map ({\bf x}) }  ,
\eqno({\rm A}2)$$
where we have canceled the constants $\beta$ in the numerator
and the denominator.  If we then invoke the identity for step
functions,
$$\Theta [a x - c] = \Theta [ x - c/a] , \eqno({\rm A}3)$$
where $a$ and $c$ are constants,
we obtain the scaling relation
$$m(\beta \map; \reduce) = m (\map; \reduce/\beta) .
\eqno({\rm A}4)$$

Similarly, using the definition [2.2] for the distribution of
volume and the scaling transformation [3.1], we can write
$$v(\beta \map; \reduce) = { \int d^n {\bf x} \,
\, \Theta \bigl[ \beta \map({\bf x}) - \reduce \bigr]
\over \int d^n {\bf x} \, \, }  =
v(\map; \reduce/\beta) , \eqno({\rm A}5)$$
where the second equality follows from applying the
identity [A3] for the step function.

Since the distributions of components $n(\map; \reduce)$ and
filaments $f(\map; \reduce)$ measure properties of the reduced
space $X^+_\reduce$, we must begin with its definition
$$X^+_\reduce \equiv \Bigl\{ {\bf x} \in D \, \Big| \,
\map ( {\bf x} ) > \reduce \Bigr\} . \eqno({\rm A}6)$$
Under the transformation $\map \to \beta \map$, the
reduced space becomes
$$\eqalign{
X^+_\reduce \to & \Bigl\{ {\bf x} \in D \, \Big| \, \beta
\map ( {\bf x} ) > \reduce \Bigr\}  \cr
\, = & \Bigl\{ {\bf x} \in D \, \Big| \, \map ( {\bf x} ) >
\reduce/\beta \Bigr\} \, . }  \eqno({\rm A}7)$$
We thus have the result
$$X^+_\reduce (\beta \map ) = X^+_{\reduce / \beta} (\map)
\, . \eqno({\rm A}8)$$
Since both output functions $n(\map; \reduce)$ and
$f(\map; \reduce)$ depend only on the properties of
the reduced space, the relation [A8] implies that
$$n(\beta \map; \reduce) = n(\map; \reduce/\beta) ,
\eqno({\rm A}9)$$
$$f(\beta \map; \reduce) = f(\map; \reduce/\beta) ,
\eqno({\rm A}10)$$
which concludes the proof. $\bullet$

\bigskip
\bigskip
\centerline{\bf APPENDIX B: PROOF OF THEOREM 2}
\medskip

In this Appendix, we present a proof of Theorem 2, which
describes how the output functions transform when the maps are
rescaled as in the nonlinear transformation of equation [3.3].

\noindent
{\it Proof:} To consider the effect of the transformation on the
distribution of density function $m(\map; \reduce)$, we use the
interpretation that $-dm/d\reduce$ is the probability that the
map has a value $\reduce$.   Thus, the probability that the
transformed map has a value of some threshold
$\reduce$ is the probability that
$$F(\map) \map = \reduce , \eqno({\rm B}1)$$
which is simply the probability that $\map = \sigwig$,
which, in turn, is given by
$$ - {d m \over d \reduce} \Bigl( \sigwig \Bigr) . \eqno({\rm B}2)$$
The transformed output function $m_T$ can then be written in
terms of the untransformed output function through
$$m_T = - \int_\reduce^\infty d\reduce \, \,
{d m \over d \reduce} \Bigl( \sigwig \Bigr) . \eqno({\rm B}3)$$
If we evaluate the integral and use the fact that $m$ vanishes
as $\reduce \to \infty$, then we find
$$m_T (\map; \reduce) = m(\map; \sigwig) . \eqno({\rm B}4)$$
Using a similar argument, we can write the transformed output
function $v_T$ for the distribution of volume in terms of the
original untransformed one via
$$v_T (\map; \reduce) = v(\map; \sigwig) . \eqno({\rm B}5)$$

Since the distributions of components $n(\map; \reduce)$ and
filaments $f(\map; \reduce)$ measure properties of the reduced
space $X^+_\reduce$, we must consider how the reduced space
transforms under equation [3.3].  This argument thus parallels
that of Appendix A.  Under the transformation of equation [3.3],
the reduced space becomes
$$\eqalign{
X^+_\reduce \to & \Bigl\{ {\bf x} \in D \, \Big| \,
F(\map) \, \map > \reduce \Bigr\}  \cr
\, = & \Bigl\{ {\bf x} \in D \, \Big| \, \map ( {\bf x} ) >
\sigwig \Bigr\} \, , }  \eqno({\rm B}6)$$
where we have used the definition of $\sigwig$.
We thus have the result
$$X^+_\reduce \bigl( F(\map) \map \bigr) = X^+_{\sigwig} (\map)
\, . \eqno({\rm B}7)$$
Since both output functions $n(\map; \reduce)$ and
$f(\map; \reduce)$ depend only on the properties of
the reduced space, the relation [B7] implies that
$$n( F(\map) \map; \reduce) = n(\map; \sigwig) ,
\eqno({\rm B}8)$$
$$f( F(\map) \map; \reduce) = f(\map; \sigwig) ,
\eqno({\rm B}9)$$
which concludes the proof. $\bullet$

\bigskip
\bigskip
\centerline{\bf APPENDIX C: PROOF OF THEOREM 3}
\medskip

In this Appendix, we present a proof of Theorem 3, which states
that the output functions $m(\map; \reduce)$, $v(\map; \reduce)$,
and $n(\map; \reduce)$ are invariant under general affine
transformations, whereas the distribution of filaments
$f(\map; \reduce)$ is invariant under similitudes.

\noindent{\it Proof:}
We begin by noting that the integral of a function $F$ transforms
under an affine transformation according to
$$\int dx dx \, F(x,y) \to |\det A| \, \int dx dy \, F(x,y) ,
\eqno({\rm C}1)$$
where $A$ is the matrix of the transformation and where $F$
is constant under the transformation (see, e.g., Barnsley 1988).

Thus, under an affine transformation, the integrals appearing
in the definitions of the distributions of density and volume
transform according to
$$\int d^n {\bf x} \, \, \map( {\bf x}) \, \,
\Theta \bigl[ \map({\bf x}) - \reduce \bigr] \, \to \,
|\det A| \, \int d^n {\bf x} \, \, \map( {\bf x}) \, \,
\Theta \bigl[ \map({\bf x}) - \reduce \bigr] \eqno({\rm C}2)$$
$$\int d^n {\bf x} \, \, \map ({\bf x})  \, \to \, |\det A| \,
\int d^n {\bf x} \, \, \map ({\bf x})  ,  \eqno({\rm C}3)$$
$$\int d^n {\bf x} \, \,
\Theta \bigl[ \map({\bf x}) - \reduce \bigr] \, \to \,
|\det A| \, \int d^n {\bf x} \, \,
\Theta \bigl[ \map({\bf x}) - \reduce \bigr] \eqno({\rm C}4)$$
$$\int d^n {\bf x} \, \, \to \, |\det A| \,
\int d^n {\bf x} \, \, . \eqno({\rm C}5)$$
Since the distribution of density $m(\map; \reduce)$ is the ratio of
the integral appearing in equation [C2] to that in equation [C3],
$m(\map; \reduce)$ is invariant under the transformation as long as
$|\det A| \ne 0$.  Similarly, equations [C4 -- C5] imply that the
distribution of volume $v(\map; \reduce)$ is also invariant under
affine transformations.

To study the effects of affine transformations on the distributions
of components $n(\map; \reduce)$ and filaments $f(\map; \reduce)$,
we must first consider the effects of the transformation on the reduced
space defined by equation [2.7], i.e., we consider the transformed space
$$w \bigl( X_\reduce^+ \bigr) = \Bigl\{ w({\bf x}) \, \Big| \,
{\bf x} \in D \, , \, \, \map ( {\bf x} ) > \reduce \Bigr\} \, .
\eqno({\rm C}6)$$
If we write the original reduced space as a union of its components
$$X_\reduce^+ = \bigcup_{j=1}^{n(\map; \reduce)} U_j , \eqno({\rm C}7)$$
where each $U_j$ is a component, then we can write the transformed
space as
$$w \bigl( X_\reduce^+ \bigr) = \bigcup_{j=1}^{n(\map; \reduce)}
w \bigl( U_j \bigr) .   \eqno({\rm C}8)$$
To show that the output function $n(\map; \reduce)$ is invariant
under the transformation, we must show that the transformed sets
$w(U_j)$ are in fact the components of the transformed space.
In order for this statement to be true, we must have:
$$w \bigl( U_j \bigr) \, \, \, {\rm connected} \, \, \,
\forall j , \eqno({\rm C}9)$$
$$w \bigl( U_j \bigr) \cap w \bigl( U_k \bigr) = \emptyset
\qquad \forall j \ne k . \eqno({\rm C}10)$$
When the affine transformation $w$ is considered only on a
set $U_j$, i.e.,
$$w: U_j \to w \bigl( U_j \bigr) , \eqno({\rm C}11)$$
then $w$ is a homeomorphism.
Hence, the first required condition [C9] is automatically
satisfied (see, e.g., Kelley 1955; Conover 1975).  We now
show that the second condition [C10] is also satisfied:
Suppose [C10] were false.
Then, by supposition, an element $\bf x$ exists such that
$${\bf x} \in w \bigl( U_j \bigr) \cap w \bigl( U_k \bigr)
\eqno({\rm C}12)$$
for some $j \ne k$. Then,
because $w$ is a homeomorphism and thus has a unique inverse,
$$w^{-1} ({\bf x}) \equiv {\bf a} \in U_j , \eqno({\rm C}13)$$
and also
$${\bf a} \in U_k . \eqno({\rm C}14)$$
The intersection $U_j \cap U_k$ is thus nonempty.  This
result contradicts the fact that $U_j$ and $U_k$ are components.
Thus, by contradiction, condition [C10] must hold.  As a result,
the distribution $n(\map; \reduce)$ of components must be invariant
under affine transformations.

We now consider how the distribution of filaments $f(\map; \reduce)$
transforms under an affine transformation.  Given the definitions
[2.9 -- 2.11],  and given that the distribution of components
$n(\map; \reduce)$ is invariant under the transformation, we must
calculate how the diameter $\diam$ and the area $\cal A$ of an open
set $U_j$ transform. The area $\cal A$ is simply the integral over the
set and the transformation is given by equation [C5] above, i.e.,
$${\cal A} \Bigl[ w(U_j) \Bigr] \, = \, |\det A| \, {\cal A} (U_j) \, ,
\eqno({\rm C}15)$$
where $w$ is the affine transformation. Under this same transformation,
the diameter of a set becomes
$$\diam \Bigl[ w(U_j) \Bigr] =
{\rm max} \Bigl\{ \big| {\bf x} - {\bf y}
\big| \, \, \Big| \, \, {\bf x} , {\bf y} \in w (U_j)
\Bigr\} . \eqno({\rm C}16)$$
Let $\bf d$ = ${\bf x} - {\bf y}$ where
${\bf x}, {\bf y}$ $\in$ $w(U_j)$. Then
$${\bf d} = w({\bf x}_0) - w ({\bf y}_0) =
A ( {\bf x}_0 - {\bf y}_0 ) , \eqno({\rm C}17)$$
where ${\bf x}_0$, ${\bf y}_0$ $\in U_j$ and where $A$
is the matrix of the affine transformation.  If the transformation
$w$ is a similitude (see equation [3.6]), then it follows that
$$\big| {\bf d} \big|^2 = r^2  \big| {\bf x}_0 - {\bf y}_0
\big|^2 \, , \eqno({\rm C}18)$$
where $r$ is the scaling factor of the transformation.
Since equation [C18] holds for all pairs of points, it follows that
$$\diam^2 \Bigl[ w(U_j) \Bigr] = r^2 \diam^2 (U_j) =
\, |\det A| \, \diam^2 (U_j) , \eqno({\rm C}19)$$
a result which holds for all similitudes.  Taken together,
equations [C15] and [C19] imply that the factor
${\cal F}_j$  = $\pi \diam_j^2 / 4 {\cal A}$
(see equation [2.10]) is invariant under similitudes.  This
result, in conjunction with the invariance of $n(\map; \reduce)$,
implies that the distribution $f(\map; \reduce)$ of filaments
(see equation [2.11]) is also invariant under similitudes. $\bullet$

We should remark that the distribution of filaments is {\it not}
invariant under a general affine transformation (as in equation [3.5]).
This claim is straightforward to verify and will not be discussed here.

\bigskip
\bigskip
\centerline{\bf APPENDIX D: RESOLUTION TRANSFORMATIONS}
\medskip

In this Appendix we derive an asymptotic expansion for the
resolution transformation $\cal R$ given by
$${\cal R} [\map ({\bf x})] = {\cal N} \int d^2 {\bf x}^\prime  \,
\exp[ - ({\bf x}^\prime - {\bf x})^2/L^2 ]
\map({\bf x}^\prime) \, , \eqno({\rm D}1)$$
where all of the quantities are the same as those defined in the
text. With the definitions
$$\xi = (x^\prime - x)/R , $$
$$\eta = (y^\prime - y)/R , \eqno({\rm D}2)$$
we can write the resolution transformation in the form
$${\cal R} [\map] = {\cal N} R^2 \int d\xi d\eta \,
\exp[ - {1 \over 2} \lambda (\xi^2 + \eta^2)  ] \,
\map(\xi, \eta) \, . \eqno({\rm D}3)$$
Integrals of the above form can be written in terms of an asymptotic
expansion when the parameter $\lambda \gg$ 1
(see Bleistein \& Handelsman 1986; see also Adams 1991).
After a considerable amount of algebra, we obtain the expansion
$${\cal R} [ \map (x,y) ] =  \sum_{j=0}^{N-1}
{1 \over (2 \lambda)^j j! } \Delta_2^j \map \Big|_{(x,y)}
+ {\cal O}\bigl( \lambda^{-N} \bigr) , \eqno({\rm D}4)$$
where $\Delta_2$ is the two dimensional Laplacian operator, i.e.,
$$\Delta_2 \equiv {\partial^2 \over \partial \xi^2} +
{\partial^2 \over \partial \eta^2} \, \, . \eqno({\rm D}5)$$
Notice that we have not written the expansion in equation [D4]
as an infinite series, but rather as the $N-1^{th}$ partial sum
with an error term $\cal O$($\lambda^{-N}$). An infinite series
is meaningless in this context because the series need not converge
(this property is a well known aspect of asymptotic analysis -- see,
e.g., Bleistein \& Handelsman 1986).

\bigskip
\bigskip
\goodbreak
\centerline{\bf APPENDIX E: ERROR ANALYSIS}
\centerline{\bf FOR THE DISTRIBUTION OF DENSITY}
\medskip

In this Appendix we estimate how uncertainties in an observed
astrophysical map will affect the output function $m(\map; \reduce)$
for the distribution of density. We also discuss how these uncertainties
affect the determination of the corresponding coordinate $\eta_m$.
Using the decomposition [4.1], the function (denoted here as $\remap$)
which appears in the numerator in equation [2.1] becomes
$$\remap = (\rmap + \delmap) \Theta \bigl[ \rmap + \delmap -
\reduce \bigr] , \eqno({\rm E}1)$$
and the integrals appearing in the definition of the mass
fraction $m$ become
$$\int d^n {\bf x} \, \remap = \int d^n {\bf x} \, \rmap \,
\Theta \bigl[ \rmap + \delmap - \reduce \bigr] +
\int d^n {\bf x} \, \delmap \,
\Theta \bigl[ \rmap + \delmap - \reduce \bigr] , \eqno({\rm E}2)$$
$$\int d^n {\bf x} \, \map = \int d^n {\bf x} \, \rmap +
\int d^n {\bf x} \, \delmap .  \eqno({\rm E}3)$$
Since the errors $\delmap$ are randomly distributed
(and hence are equally likely to be positive or negative),
the integrals of $\delmap$ tend toward zero.  Notice that the
step function introduces a different measure of integration,
but does not otherwise change this result.\footnote{$^\dagger$}
{This statement is only correct to first order in $\delmap$.
Since the $\delmap$ integral in equation [E2] contains a
step function which includes the error contribution $\delmap$,
noise in the map will push more pixels above the threshold $\reduce$
than below.  However, this effect is second order in $\delmap$.
To show this, we expand the step function in the integral
in equation [E2] to obtain $\Theta \bigl[ \rmap + \delmap -
\reduce \bigr]$ = $\Theta \bigl[ \rmap - \reduce \bigr]$
$- \delmap \, \delta \bigl[ \rmap - \reduce \bigr]$.
Using this expansion in the original integral, we see that
a $\delmap^2$ term appears. This term is positive definite
and hence does not integrate to zero; however, the term is
second order and can be ignored.}

Thus, the distribution of density $m_{\rm obs}$ that is calculated
from the observed map $\omap$ is related to the true distribution
$m_R$ through
$$m_{\rm obs} (\map ; \reduce ) = m_R (\map ; \newsig) ,
\eqno({\rm E}4)$$
where we have defined
$$\newsig \equiv \reduce - \delmap . \eqno({\rm E}5)$$
In other words, the net effect of the uncertainty in the
observed map is to introduce an uncertainty in the variable
$\reduce$ of the output function.

Using equations [E4] and [E5], we can calculate the associated
error in the output function $m(\reduce)$ for a give map
through a Taylor series expansion:
$$m_R(\reduce) = m_{\rm obs} (\newsig) =
m_{\rm obs} (\reduce) +
{d m \over d \reduce} \Big|_\reduce \delmap +
{\cal O} \bigl[ (\delmap)^2 \bigr] . \eqno({\rm E}6)$$
To leading order the error in the output function is then
approximately given by
$$ {\sl error} = \Delta m \sim \Big| {d m \over d \reduce} \Big| \,
\delmap , \eqno({\rm E}7)$$
which is first order in $\delmap$ and therefore small if the signal
to noise ratio in the original map is sufficiently high.

Given the expansion [E6] for the error in the output function,
we now calculate the corresponding uncertainty in the coordinate
$\eta_m$. Using the expansion [E6] in the definition of the
coordinate (see Paper I), we can show that
$$\langle \Sigma \rangle (\eta_{m R}^2 - \eta_{m {\rm obs}}^2 )
= - \int_0^{\map_C} 2 \, (1 - m_{\rm obs}) {dm \over d \reduce} \,
\delmap \, d\reduce \, + \, \int_{\map_C}^\infty \, 2 \, m_{\rm obs}
\, {dm \over d \reduce} \, \delmap \, d\reduce \, . \eqno({\rm E}8)$$
Since the uncertainty $\delmap$ can be either positive or negative,
the integrals in equation [E8] tend to cancel out.  If, however, we
consider the uncertainty to be bounded $|\delmap| \le \bound$,
we can place corresponding bounds on the integrals:
$$\Bigg| \int_0^{\map_C} 2 \, (1 - m_{\rm obs}) {dm \over d \reduce}
\, \delmap \, d\reduce \, \Bigg| \, \le \, \bound/4 ,
\eqno({\rm E}9{\rm a}) $$
$$\Bigg| \int_{\map_C}^\infty \, 2 \, m_{\rm obs}
{dm \over d \reduce} \, \delmap \, d\reduce \, \Bigg|
\, \le \, \bound/4 , \eqno({\rm E}9{\rm b}) $$
where we have used the fact that $m(0)=1$, $m(\map_C) = 1/2$,
and $m(\infty) = 0$.  Combining all of the above results
shows that
$$\Delta \eta_m \le {\bound \over 4 \eta_m \langle \reduce \rangle }
\, + \, {\cal O} (\bound^2) \, + \, {\cal O} (\Delta \eta^2) .
\eqno({\rm E}10)$$

\bigskip
\bigskip
\centerline{\bf APPENDIX F: ERROR ANALYSIS}
\centerline{\bf FOR THE DISTRIBUTION OF COMPONENTS}
\medskip

In this Appendix we discuss how observational errors in
the map $\map$ can produce errors in the output function
$n(\map; \Sigma)$ which determines the number of topological
components as a function of threshold density $\reduce$.

As discussed in the text (\S 4),
we must estimate the probability that a given pixel
in a map will have an observed value $\reduce_{\rm obs}$
which is greater than a given threshold value $\reduce_0$ when
in fact the true value $\reduce_T$ of the pixel is less than
that of the threshold. We first assume that the probability
of error in the observed value has a Gaussian distribution
so that the probability of having an observed value
$\reduce_{\rm obs}$ and a true value $\reduce_T$ is given by
$$p(\reduce_{\rm obs}, \reduce_T)  =
{\cal N} \exp \bigl[ - (\reduce_{\rm obs} -
\reduce_T)^2 / \std^2 \bigr] , \eqno({\rm F}1)$$
where $\cal N$ is the normalization factor and where $\std$ is the
standard deviation.  For a given threshold $\reduce_0$, the probability
$P_1 (\reduce_0)$ that an observed value is greater than the threshold
and that the true value is less than the threshold is given by
$$P_1 (\reduce_0) = {\cal N} \int_{\reduce_0}^\infty \, d\reduce_{\rm obs}
\, \Bigl[ - {d m \over d \reduce} \Bigr]_{\rm obs} \,
\int_0^{\reduce_0} \, d\reduce_T \, \exp
\bigl[ - (\reduce_{\rm obs} - \reduce_T)^2 / \std^2 \bigr] ,
\eqno({\rm F}2)$$
where we have integrated over all possible values $\reduce_{\rm obs}$
$\ge \reduce_0$ and all possible values $\reduce_T \le \reduce_0$
and we have used the probability $-dm/d\reduce$ that the
observed value is $\reduce_{\rm obs}$.

Although the probability $P_1$ depends on the threshold $\reduce_0$
and on the distribution of density $m(\map; \reduce)$ for the map,
we can derive a useful upper bound.  We first introduce non-dimensional
quantities
$$x = (\reduce_{\rm obs} - \reduce_T)/\std , $$
$$a = (\reduce_{\rm obs} - \reduce_0)/\std \ge 0 , \eqno({\rm F}3)$$
$$b = \reduce_{\rm obs}/\std \ge a . $$
The integral over $\reduce_T$ appearing in equation [F2] can then
be written in the form
$$\std \int_a^b \, e^{-x^2} \, dx \, \le \std
{ {\sqrt \pi} \over 2 } e^{-a^2} , \eqno({\rm F}4)$$
where the final inequality is straightforward to show. Using the
bound of equation [F4] in the expression [F2] for $P_1$, we obtain
the bound
$$P_1 (\reduce_0) \le {1 \over 2} \int_{\reduce_0}^\infty \,
d\reduce_{\rm obs} \, \Bigl[ - {d m \over d \reduce} \Bigr]_{\rm obs}
\, \exp \bigl[ - (\reduce_{\rm obs} - \reduce_0)^2 / \std^2 \bigr] ,
\eqno({\rm F}5)$$
where we have used the normalization condition for the Gaussian
probability distribution (see equation [F1]), i.e.,
$$2 {\cal N} \std \, \int_0^\infty \, e^{-v^2} \, dv \,
= {\cal N} \std {\sqrt \pi} = 1 \, . \eqno({\rm F}6)$$
A weaker bound on the probability $P_1$ can be found by replacing
$-dm/d\reduce$ by its maximum value and evaluating the remaining
integral to obtain
$$P_1(\reduce_0) \le { {\sqrt \pi} \over 4 } \, \std \Bigl[
- {d m \over d \reduce} \Bigr]_{\rm max} \, \, . \eqno({\rm F}7)$$

In practice, we can use the observed distribution of density
profile $m(\map; \reduce)$ to find the maximum value of
$-dm/d\reduce$ and thus constrain $P_1$.  However, we can
obtain a crude numerical estimate for $P_1$ as follows.  We write
$$- \std {dm \over d\reduce} \approx {\std \over \Delta \reduce} \, ,
\eqno({\rm F}8)$$
where $\Delta \reduce$ is the overall dynamic range of the map.
The ratio in equation [F8] is thus the inverse of the average
signal to noise ratio of the map.  We thus obtain the estimate
$$P_1 \approx {\sqrt \pi \over 4} {\std \over \Delta \reduce}
\approx 10^{-3}, 10^{-2} \, \, ,  \eqno({\rm F}9)$$
where we haved used signal to noise ratios of $\sim$ 400
and $\sim$ 40 to obtain the numerical estimates.  These values
are representative of the {\it IRAS} maps considered in our
companion paper (WA).

Thus, for a map with a large number of pixels, many
pixels can be ``wrong'' at a given threshold in the sense
that the pixel value appears to be larger than the
threshold value when in fact the true value of less
than the threshold. We note that this estimate represents
an upper bound on the probability $P_1$ and we also note
that only a small fraction of these ``erroneous'' pixels will
produce spurious topological components.  As we discuss in the
text, the number of spurious components (those components produced
by pixels erroneously appearing larger than a threshold) can be
controlled if we restrict our attention to components with three
or more pixels (see \S 4).

\bigskip
\bigskip
\centerline{\bf APPENDIX G: ERROR ANALYSIS}
\centerline{\bf FOR THE DISTRIBUTION OF FILAMENTS}
\medskip

In this Appendix, we consider the effects of observational
errors on the output function $f(\map; \reduce)$ which measures
the filamentary nature of the map.  As defined in \S 2, the
filament function is a sum over all of the components of the
map.  For a given threshold level $\reduce$, we separate
the true components $n_R$ from the spurious ones $\Delta n$ through
$$n_{\rm obs} = n_R + \Delta n , \eqno({\rm G}1)$$
where it is understood that all three quantities are
functions of the threshold level $\reduce$.  The filament
function can then be written in the form
$$f_{\rm obs} = {1 \over n_R + \Delta n } \sum_{j=0}^{n_R}
( {\cal F}_j + \Delta {\cal F}_j ) +
{1 \over n_R + \Delta n } \sum_{k=0}^{\Delta n}
( {\cal F}_k + \Delta {\cal F}_k ) , \eqno({\rm G}2)$$
where $\Delta {\cal F}_j$ is the uncertainty in the filament
index of a given component.  If we let $f_R$ denote the
true value of the distribution of filaments, the uncertainty
can be written in the form
$$\Delta f = f_R - f_{\rm obs} =
{\Delta n \over n_R + \Delta n }
\Bigl\{ f_R - {1 \over \Delta n}
\sum_{k=0}^{\Delta n} {\cal F}_k \Bigr\}
- {1 \over n_R + \Delta n } \sum_{j=0}^{n_R + \Delta n}
\Delta {\cal F}_j , \eqno({\rm G}3)$$
where the first term in brackets is due to the uncertainty in
the number of components, the second term in brackets is the
contribution of the spurious components, and the final sum is the
error due to the uncertainty in each individual filament index.
Since the uncertainties $\Delta {\cal F}_j$ can be either
positive or negative, this final sum will tend toward zero.
Let us therefore concentrate on the terms in brackets.  The
sum in brackets is the filament index of the spurious components
and we will denote this quantity as $f_S$.  The relative error
in the distribution of filaments can then be written
$${\Delta f \over f_R} = {\Delta n \over n_R + \Delta n }
\Bigl\{ 1 - {f_S \over f_R} \Bigr\} \approx
{\Delta n \over n_R} \Bigl\{ 1 - {f_S \over f_R} \Bigr\}
\, , \eqno({\rm G}4)$$
where we have assumed that $\Delta n$ $\ll$ $n_R$ in obtaining
the second approximate equality.  As we argue in the text (see
\S 4), the term in brackets is expected to be of order unity.
Thus, the relative error in the filament index is small if the
relative error in the distribution of components can also be
made small.

\bigskip
\bigskip
\centerline{\bf APPENDIX H: RELATIONSHIP BETWEEN THE}
\centerline{\bf DISTRIBUTIONS OF DENSITY AND VOLUME}
\medskip

In this appendix, we present a simple relationship between the
distribution of density $m(\map; \reduce)$ and the distribution
of volume $v(\map; \reduce)$.   For any map $\map$ and for any
threshold level $\reduce$,  the distribution of density
$m(\map; \reduce)$ is always greater than the corresponding
distribution of volume $v(\map; \reduce)$.  We show that this
claim is true by the following argument:

We begin by defining an average density $\mapbar$,
$$\mapbar \, = {1 \over A_n} \, \int d^n {\bf x} \, \,
\map ({\bf x}) \qquad {\rm where} \qquad A_n =
\int d^n {\bf x} \, \, ,  \eqno({\rm H}1)$$
and where the integrals are taken over the entire map.
The subscript $n$ denotes the number of spatial dimensions
(this argument holds for arbitrary $n$).
The distribution of density can now be written in the form
$$m(\map; \reduce) =  {1 \over A_n} \, \int d^n {\bf x} \, \,
{\map({\bf x}) \over \mapbar} \Theta [ \map - \reduce ] .
\eqno({\rm H}2)$$
The distribution of volume can be written in the form
$$v(\map; \reduce) = {1 \over A_n} \, \int d^n {\bf x} \, \,
\Theta [ \map - \reduce ] \, . \eqno({\rm H}3)$$
For high threshold levels, i.e., $\reduce > \mapbar$,
the quantity $\map({\bf x})/\mapbar$ is greater than
unity whenever the integrand of either equation [H2] or [H3]
is nonzero. It follows directly that
$$m(\map; \reduce) > v(\map; \reduce) \qquad {\rm for} \qquad
\reduce > \mapbar . \eqno({\rm H}4)$$
We must now consider the case when the threshold level
$\reduce < \mapbar$.  The derivatives of the two distributions
(with respect to the threshold $\reduce$) are given by
$$- {dm \over d\reduce} = {1 \over A_n} \,
\int d^n {\bf x} \, \, {\map({\bf x}) \over \mapbar}
\delta [ \map - \reduce ] \, , \eqno({\rm H}5)$$
$$- {dv \over d\reduce} = {1 \over A_n} \, \int d^n {\bf x}
\, \, \delta [ \map - \reduce ] \, . \eqno({\rm H}6)$$
For the threshold range of interest, $\reduce < \mapbar$,
the quantity $\map({\bf x})/\mapbar$ is less than
unity whenever the integrand of either equation [H5] or [H6]
is nonzero (i.e., whenever $\map = \reduce$).  As a result,
$$\Bigg| {dm \over d\reduce} \Bigg| <
\Bigg| {dv \over d\reduce} \Bigg| \eqno({\rm H}7)$$
for the threshold range of interest. Since both functions start at
unity, $m(\map; 0) = 1 = v(\map; 0)$, and both function have negative
slopes, equation [H7] implies that
$$m(\map; \reduce) > v(\map; \reduce)
\qquad {\rm for} \qquad \reduce < \mapbar .
\eqno({\rm H}8)$$
Thus, by equations [H4] and [H8], the distribution of density
$m(\map; \reduce)$ is always greater than the distribution of
volume $v(\map; \reduce)$.

\newpage
\bigskip
\bigskip
\centerline{\bf REFERENCES}
\medskip

\par\pp
Adams, F. C. 1991, {\sl ApJ}, {\bf 382}, 544

\par\pp
Adams, F. C. 1992, {\sl ApJ}, {\bf 387}, 572 (Paper I)

\par\pp
Ballard, D. H., \& Brown, C. M. 1982, {Computer Vision}
(Englewood Cliffs, New Jersey: Prentice-Hall)

\par\pp
Barnsley, M. F. 1988, {Fractals Everywhere}
(San Diego: Academic Press)

\par\pp
Bazell, D., \& D\'esert, F. X. 1988, {\sl ApJ}, {\bf 333}, 353

\par\pp
Bleistein, N. \& Handelsman, R. A. 1986, {Asymptotic Expansions
of Integrals} (New York: Dover)

\par\pp
Conover, R. A. 1975, {A First Course in Topology}
(Baltimore: Williams \& Wilkins)

\par\pp
Copson, E. T. 1968, {Metric Spaces}
(London: Cambridge University Press)

\par\pp
Dickman, R. L., Horvath, M. A., \& Margulis, M. 1990,
{\sl ApJ}, {\bf 365}, 586

\par\pp
Edgar, G. A. 1990, Measure, Topology, and Fractal Geometry
(New York: Springer Verlag)

\par\pp
Elizalde, E. 1987, {\sl J. Math. Phys.}, {\bf 28} (12), 2977

\par\pp
Elmegreen, B. G. 1993, in Protostars and Planets III,
ed. E. Levy and M. S. Mathews (Tucson: University of
Arizona Press), p. 97

\par\pp
Falgarone, E., Phillips, T. G., \& Walker, C. K. 1991,
{\sl ApJ}, {\bf 378}, 186

\par\pp
Gott, J. R., Melott, A. L., \& Dickinson, M. 1986,
{\sl ApJ}, {\bf 306}, 341

\par\pp
Houlahan, P., \& Scalo, J. M. 1992, {\sl ApJ}, {\bf 393}, 172

\par\pp
Jarrett, T. H., Dickman, R. L., \& Herbst W. 1989,
{\bf 345}, 881

\par\pp
Kelley, J. L. 1955, {General Topology} (Princeton: D. Van Nostrand)

\par\pp
Langer, W. D., Wilson, R. W., Goldsmith, P., \& Beichman, C. A.
1989, {\sl ApJ}, {\bf 337}, 355

\par\pp
Larson, R. B. 1981, {\sl MNRAS}, {\bf 194}, 809

\par\pp
Larson, R. B. 1985, {\sl MNRAS}, {\bf 214}, 379

\par\pp
Lord, E. A., \& Wilson, C. B. 1984, {The Mathematical
Description of Shape and Form} (Sussex: Ellis Horwood)

\par\pp
Melott, A. L. 1990, {\sl Phys. Rep.}, {\bf 193}, 1

\par\pp
Myers, P. C. 1991, in {Fragmentation of Molecular Clouds
and Star Formation} (IAU Symposium 147), eds. E. Falgarone,
F. Boulanger, and G. Duvert (Dordrecht: Kluwer), p. 221

\par\pp
Scalo, J. 1990, in {Physical Processes in Fragmentation
and Star Formation}, eds. R. Capuzzo-Dolcetta et al.
(Dordrecht: Kluwer), p. 151

\par\pp
Shandarin, S. F., \& Zel'dovich Ya. B. 1989, {\sl Rev. Mod. Phys.},
{\bf 61}, 185

\par\pp
Veeraraghavan, S., \& Fuller, G. A. 1991, in
{Fragmentation of Molecular Clouds and Star Formation}
(IAU Symposium 147), eds. E. Falgarone, F. Boulanger,
and G. Duvert (Dordrecht: Kluwer), p. 221

\par\pp
Williams, J., de Geus, E., \& Blitz, L. 1994, {\sl ApJ}, in press

\par\pp
Wiseman, J. J., \& Adams, F. C. 1994, submitted to {\sl ApJ} (WA)

\par\pp
Wood, D. O. S., Myers, P. C., \& Daugherty, D. A. 1994,
{\sl ApJ Suppl.}, in press

\bye